\documentclass[fleqn,usenatbib]{mnras}

\usepackage{newtxtext,newtxmath}

\usepackage[T1]{fontenc}

\DeclareRobustCommand{\VAN}[3]{#2}
\let\VANthebibliography\thebibliography
\def\thebibliography{\DeclareRobustCommand{\VAN}[3]{##3}\VANthebibliography}

\usepackage{graphicx}	
\usepackage{amsmath}	

\usepackage{graphicx}

\usepackage{multirow}

\usepackage{longtable}
\usepackage{enumitem}

\usepackage{soul}

\usepackage{dcolumn}
\newcolumntype{d}[1]{D{.}{.}{#1}}

\usepackage{siunitx}

\usepackage[]{xcolor}

\definecolor{tbc}{cmyk}{0.05002,1,0.9,0}

\definecolor{mygreen}{cmyk}{0.65002,0.1,1,0}

\definecolor{myyellow}{cmyk}{0.00,0.24,0.86,0.20}

\definecolor{mypink}{cmyk}{0.00,1.00,0.00,0.00}

\definecolor{tbud}{cmyk}{0.1002,0.0,0,0.4}

\definecolor{mygrey}{cmyk}{0.180,0.180,0.180,0.180}

\newcommand{\kmps}{$\rm km\,s^{-1}$}

\newcommand{\lum}{$\rm erg\,s^{-1}$}

\newcommand{\obj}{\mbox{Mrk\,335}}

\newcommand{\hst}{\mbox{{\it HST}}}
\newcommand{\gaia}{\mbox{{\it Gaia}}}

\title[Parsec-scale jet in \obj]{Detection of a Parsec-Scale Jet in a Radio-Quiet Narrow-Line Seyfert 1 Galaxy with Highly Accreting Supermassive Black Hole}

\author[S. Yao et al.]{
Su Yao,$^{1}$\thanks{E-mail: syao@mpifr-bonn.mpg.de}
Xiaolong Yang,$^{2,3}$\thanks{E-mail: yangxl@shao.ac.cn}
Minfeng Gu,$^{4}$
Tao An,$^{2}$
Jun Yang,$^{5}$
Luis C. Ho,$^{3,6}$
Xiang Liu,$^{7}$
Ran Wang,$^{3}$
\newauthor
Xue-Bing Wu,$^{3,6}$
Weimin Yuan$^{8}$
\\
$^{1}$Max-Planck-Institut f\"ur Radioastronomie, Auf dem H{\"u}gel 69, 53121 Bonn, Germany\\
$^{2}$Key Laboratory of Radio Astronomy, Shanghai Astronomical Observatory, Chinese Academy of Sciences, Shanghai 200030, China\\
$^{3}$Kavli Institute for Astronomy and Astrophysics, Peking University, Beijing 100871, China \\
$^{4}$Key Laboratory for Research in Galaxies and Cosmology, Shanghai Astronomical Observatory, Chinese Academy of Sciences, Shanghai 200030, China\\
$^{5}$Department of Space, Earth and Environment, Chalmers University of Technology, Onsala Space Observatory, SE-439 92 Onsala, Sweden\\
$^{6}$Department of Astronomy, School of Physics, Peking University, Beijing 100871, China\\
$^{7}$Key Laboratory of Radio Astronomy, Xinjiang Astronomical Observatory, Chinese Academy of
Sciences, 150 Science 1-Street, Urumqi 830011, China\\
$^{8}$Key Laboratory of Space Astronomy and Technology, National Astronomical Observatories, Chinese Academy of Sciences, Beijing 100101, China
}

\date{Accepted XXX. Received 2021; in original form}

\pubyear{2021}

\begin{document}
\label{firstpage}
\pagerange{\pageref{firstpage}--\pageref{lastpage}}
\maketitle

\begin{abstract}
The jet in active galactic nuclei (AGN) is a key ingredient in understanding the co-evolution of galaxies and their central supermassive black holes (SMBHs). 
Unfortunately, the mechanism of jet launching and collimation is still elusive. 
The observational evidence of decreasing radio loudness with increasing Eddington ratio implies that the jet should be coupled with the accretion process. 
To further explore the relationship between the jet and accretion, it is necessary to extend our knowledge of the jet to an extreme end of the Eddington ratio distribution of AGN. Using Very Long Baseline Array (VLBA), we report the detection of the parsec-scale radio structure in \obj, a radio-quiet narrow-line Seyfert 1 galaxy with an Eddington ratio close to/above unity. 
The VLBA image at 1.5\,GHz reveals an elongated structure extending $\sim20$\,parsec in north-south direction with a peak flux density of $1.98\pm0.05$\,mJy/beam and radio brightness temperatures as high as 
$6\times10^{7}$\,K. 
This feature provides a strong evidence of a parsec-scale (bipolar) jet launched from a highly accreting SMBH. We discuss the result by comparing \obj\ with other highly accreting systems, e.g. Galactic black holes and tidal disruption events, 
and recall the discovery of collimated corona in the vicinity  
of SMBH in \obj\ by previous X-ray observations, 
whose relation to the parsec-scale radio jet should be explored by future simultaneous X-ray spectroscopy and high resolution radio observations. 
%
\end{abstract}




\section{Introduction}\label{sec:intro}

The collimated outflowing plasma, 
also termed as `jet', 
launched from active galactic nuclei (AGN) 
is suggested to be one of the important forms of AGN feedback 
which regulates the coevolution of supermassive black hole (SMBH) and their host galaxy 
\citep[][]{2012ARA&A..50..455F, 2013ARA&A..51..511K}. 
The AGN jets typically peak their radiation in the radio band by synchrotron processes and could be luminous radio sources. 
However, only 15--20 per cent of the AGN are radio-loud \citep[][]{1995PASP..107..803U, 2002AJ....124.2364I}, 
indicating that not all of the AGN launch jets. 
This raises the questions of why we only observe jets in a small fraction of AGN 
and how the jets are formed. 

The observational evidence of a strong anti-correlation between the radio loudness parameter $\mathcal{R}$\footnote{%
The radio loudness parameter $\mathcal{R}$ is defined as the ratio of radio-to-optical flux density, conventionally at 5\,GHz and $B$-band (4400\AA), respectively, 
with $\mathcal{R}\approx10$ as the dividing of radio-quiet and radio-loud AGN \citep[e.g.][]{1989AJ.....98.1195K}. }
and the bolometric luminosity expressed in Eddington units, i.e. Eddington ratios $\lambda_{\rm Edd}$, 
implies coupling of the jet and accretion process 
\citep[][]{2002ApJ...564..120H, 2008ARA&A..46..475H, 2006ApJ...636...56G, 2007ApJ...658..815S, 2020ApJ...904..200Y}; 
the launch of jet may be dependent on the Eddington ratio 
and underlying physics 
such as the mass accretion rate, the state of accretion disk and the growth of SMBH. 
But the detailed mechanism 
remains elusive. 
Given the anti-correlation of the radio loudness and the Eddington ratio, 
questions arise as to
whether the collimation of outflows is suppressed as the Eddington ratio increases, 
and 
whether there is a limiting Eddington ratio above which the collimation is quenched. 
On the other hand, 
while the origin of radio core emission is clear in the radio-loud AGN as dominated by the non-thermal radiation from jets, 
the case of the radio-quiet AGN is more complex. 
With milli-arcsecond (mas) resolution radio images 
achieved by very-long-baseline interferometry (VLBI), 
it was found that some of the radio-quiet AGN reveal compact parsec-scale nuclear radio emission with brightness temperatures of $T_{\rm B}\gtrsim10^{7}\rm\,K$ and occasionally accompanied by collimated features 
which provide the evidence of small, weak jets 
\citep[e.g.][]{2003ApJ...583..192M, 2004A&A...417..925M, 2009ApJ...706L.260G, 2013MNRAS.432.1138P, 2013ApJ...765...69D, 2021MNRAS.tmp..609W}, 
while others only show radio emission from the presumably star forming region, disk wind or hot corona
\citep[e.g.][see also \citealt{2019NatAs...3..387P} for a review]{2004ApJ...613..794G, 2012MNRAS.426..588B}.
Questions arise as to why the collimated outflows do form in some radio-quiet AGN but not in others, 
and whether they are intrinsically different from 
their powerful analogs in blazars and radio galaxies.

Keeping in mind the above questions, 
we have proposed imaging observations 
on the central parsec-scale nuclear region of a sample of nearby radio-quiet narrow-line Seyfert 1 (NLS1) galaxies with accretion rates approaching to or exceeding the Eddington limit \citep[][]{2020ApJ...904..200Y} 
using the Very Long Baseline Array (VLBA) 
of the National Radio Astronomy Observatory (NRAO). 
The NLS1 galaxies are believed to rapidly growing their central SMBHs that are in a relatively low mass range of $\sim10^{6}-10^{8}\rm\,M_{\odot}$ \citep[][]{2006ApJS..166..128Z}. 
They are often being radio-quiet and have a lower radio-loud fraction compared to the normal Seyfert galaxies and quasars \citep[][]{2008ApJ...685..801Y}. 
While the parsec-scale jets have been detected in the radio-loud NLS1 galaxies 
\citep[e.g.][]{2011ApJ...738..126D, 2015ApJS..221....3G}, 
the case is still elusive for the radio-quiet ones, 
especially when it approaches the highest Eddington ratio regime. 
In recent years, 
the work dedicated to study parsec-scale region of the radio-quiet NLS1 galaxies has found jet features in several of them \citep[][]{2013ApJ...765...69D}, 
but only Mrk~1239 has a reliably measured Eddington ratio close to/higher than one\footnote{%
Another one is Ark~564, the VLBI image of which shows an elongated structure as reported by \citet{2004A&A...425...99L}. 
However, it is noted in their paper that the ($u$, $v$) coverage obtained was relatively poor, 
and they do not feel confident that the elongated feature is an accurate representation of the VLBI-scale structure of the source.} 
\citep[][]{2015ApJ...798L..30D, 2021ApJ...912..118P}. 
So further studies on the parsec-scale nuclear region in radio-quiet NLS1 galaxies with highest Eddington ratios should provide general clues of the interplay between the jet and accretion disk. 
Our observations will not only search the signature of jet in these radio-quiet highly accreting systems, 
but also extend our knowledge of the parsec-scale radio properties of AGN to a rarely explored parameter space.

In this paper, 
we present the result of our VLBA observation on \obj\ \citep[$z=0.025785$;][]{1999ApJS..121..287H} as one of the outcomes in the course of our ongoing project (Yang et al. in preparation). 
\obj\ is a well-known NLS1 galaxy with a SMBH weighing $\sim10^{7}\rm\,M_{\odot}$ \citep[][]{2004ApJ...613..682P, 2012ApJ...744L...4G, 2014ApJ...782...45D} 
hosted in a disk galaxy \citep[][]{2017ApJS..232...21K}. 
It has been extensively observed in optical \citep[e.g.][]{2004ApJ...613..682P, 2012ApJ...744L...4G, 2014ApJ...782...45D} 
and X-rays \citep[e.g.][]{2008ApJ...681..982G, 2018MNRAS.478.2557G, 2019MNRAS.484.4287G, 2020MNRAS.499.1266T, 2020A&A...643L...7K}. 
Based on its optical luminosity and hard X-ray spectrum, 
\obj\ was suggested to host a candidate of super-Eddington accreting massive black hole \citep[][]{2013PhRvL.110h1301W,2014ApJ...793..108W}.
But the knowledge of its radio emission is rather limited. 
\obj\ has been observed by the VLA with various angular resolution 
(see Table~\ref{tab:vla_flux} and the literatures therein). 
All these observations reveal a compact unresolved weak source. 
Based on the nuclear optical flux derived from the decomposition of HST $V$-band (F550M) image 
\citep[][]{2014ApJ...782...45D} and an assumed optical spectral index of $\alpha=-0.5$ 
($S_{\nu}\propto\nu^{\alpha}$), 
the radio loudness of \obj\ is estimated to be $\mathcal{R}\approx0.8$, 
falling into the radio-quiet category. 
Interestingly, a collimated outflow near the accretion disk has been discovered in \obj\ from the X-ray spectral analysis by \citet{2019MNRAS.484.4287G}. 
Here we report the detection of a parsec-scale jet feature in \obj\ using VLBA. 
Throughout this work a cosmology is assumed with $H_0=70$ km s$^{-1}$ Mpc$^{-1}$, $\Omega_\Lambda=0.73$ and $\Omega_{\rm M}=0.27$; 
1\,mas corresponds to a scale of 0.512\,parsec at the distance of \obj.





\begin{table}
\setlength{\tabcolsep}{4pt}
\footnotesize
    \caption{Previous VLA observations on \obj.
    }
    \label{tab:vla_flux}
    \centering                          
    \begin{tabular}{c c c c l}        
    \hline\hline                 
    \multicolumn{1}{c}{Frequency} & 
    \multicolumn{1}{c}{Flux density} & 
    \multicolumn{1}{c}{Date} & 
    \multicolumn{1}{c}{Resolution} &
    \multicolumn{1}{c}{Reference} \\
    \multicolumn{1}{c}{[GHz]} & 
    \multicolumn{1}{c}{[mJy]} & 
    \multicolumn{1}{c}{} & 
    \multicolumn{1}{c}{} & 
    \multicolumn{1}{c}{} \\
    \multicolumn{1}{c}{(1)} & 
    \multicolumn{1}{c}{(2)} & 
    \multicolumn{1}{c}{(3)} & 
    \multicolumn{1}{c}{(4)} & 
    \multicolumn{1}{c}{(5)} \\ 
    \hline                        
    \noalign{\smallskip}
    1.4 & 4.1$\pm$0.9 & 1983.07 & 1$\farcm$5 & E87 \\
      & 6.19$\pm$0.29 & 1992.12 & 2$\arcsec$ & N99,Y20 \\
      & 7.3$\pm$0.4 & 1993.11 & 45$\arcsec$ & C98 \\
    \noalign{\smallskip}
     5 & 3.9 & 1982.11-1983.05 & 18$\arcsec$ & K89,M93 \\
      & 3.3$\pm$0.2 & 1983.07 & 15$\arcsec$ & E87 \\
      & 3.58$\pm$0.05 & 2003.06$^{\rm*}$ & 0$\farcs$3 & L06 \\
      & 3.21$\pm$0.02 & 2015.07 & 0$\farcs$4 & B18,Y20 \\
    \noalign{\smallskip}
     8.4 & 2.05 & 1991.06 & 0$\farcs$3 & K95 \\ 
      & 2.23 & 1992.04 & 2$\farcs$5 & K95 \\
      & 2.14$\pm$0.06 & 1992.12 & 0$\farcs$3 & N99,Y20 \\
      & 2.3$^{\ddagger}$ & 1997.01-1999.04 & 1$\arcsec$--3$\arcsec$ & B05 \\
    \noalign{\smallskip}
    \hline                                   
\end{tabular}
\parbox[]{\columnwidth}{%
    {\bf Note. }
    E87: \citet{1987ApJ...313..651E}; N99: \citet{1999ApJS..120..209N}; C98: \citet{1998AJ....115.1693C}; 
    K89: \citet{1989AJ.....98.1195K}; M93: \citet{1993MNRAS.263..425M}; L06: \citet{2006A&A...455..161L}; 
    B18: \citet{2018A&A...614A..87B}; Y20: \citet{2020ApJ...904..200Y}; K95: \citet{1995MNRAS.276.1262K}; 
    B05: \citet{2005ApJ...618..108B}. 
    \\
    $^{*}$ This observation date is obtained from the VLA data archive. \\
    $^{\ddagger}$ This measurement is the mean flux density obtained from a number of observations (see B05).
    }
\end{table}


\section{Observations and Data Reduction} \label{sec:observation}

\subsection{VLBA Observation and Data Reduction} \label{sec:vlba_observation}
We observed \obj\ on September 15, 2018 using 10 antennas of the VLBA (project code: BY145, P.I. Yao). 
The observation was made in the L-band with a central frequency of 1.548\,GHz ($\lambda\sim19.4\rm\,cm$), and a total bandwidth of 256\,MHz. 
The observation was performed in the phase-referencing mode: the target field was observed for 1.53 hours in 4-minute scans, 
interleaved with 1-minute scans of the phase referencing calibrator J0004$+$2019 (RA: 00$^\mathrm{h}$04$^\mathrm{m}$35$^\mathrm{s}.758287$, Dec: $+$20$^\circ$19$^\prime$42$\farcs$31786, J2000), with the phase calibrator located at 0.42\,deg away from the target. We also inserted a scan of the bright radio source 3C\,454.3 for fringe and bandpass calibration. The data were recorded at 2 Gbps in dual circular polarization. The raw data was correlated with the software correlator DiFX \citep{2007PASP..119..318D, 2011PASP..123..275D}. 

The visibility data was calibrated using the NRAO Astronomical Image Processing System \citep[{\sc aips},][]{2003ASSL..285..109G} following the standard procedure. 
A prior amplitude calibration was performed with the system temperatures and the antenna gain curves. 
The earth orientation parameters were obtained and corrected by using the measurement from the U.S. Naval Observatory data base and the ionospheric dispersive delays were corrected according to a map of total electron content provided by the GPS satellite observations.
The opacity and parallactic angle corrections were also applied according to the attached information in data. 
Phase delay from the instrument was removed by fringe-fitting over the scan of 3C\,454.3. 
The bandpass solutions were also determined from the fringe finder calibrator 3C\,454.3. 
Finally, we performed a global fringe-fitting on the phase-referencing calibrator J0004$+$2019. 
The calibrated data of the phase-referencing source J0004$+$2019 was averaged in {\sc aips} and then exported into {\sc difmap} software package \citep{1997ASPC..125...77S} for imaging and self-calibration. 
J0004$+$2019 shows a bright and compact structure in our observation, consistent with the model assumed in fringe-fitting in {\sc aips}.

The deconvolution of the visibility data was performed in {\sc difmap}. 
The source is relatively weak, with a peak flux density of only $1.98\pm0.05\,\mathrm{mJy/peak}$ and a signal-to-noise ratio of $\sim47$. 
Therefore, we did not perform the self-calibration procedure. 

We also check the historical VLBA observations of \obj\ in projects BB0056 and BG0093 from the NRAO archive. 
The observation in project BB0056 is performed in the X-band on June 9, 1996 
with a total bandwidth of 32\,MHz and a total on-source time of 48 minutes, 
while the observation in project BG0093 is performed in the S/X-band%
\footnote{
We note that the observing band of BG0093 in the NRAO archive is wrongly marked as L-band.}
on July 9, 1999 
with a total bandwidth of 192\,MHz and also a total on-source time of 48 minutes. 
We have processed all these data following similar procedures as described above. 
Unfortunately, \obj\ is not detected in the X-band observation during BB0056 and BG0093. 
This is consistent with the result reported by \citet[][]{1998MNRAS.299..165B}, 
which also analyzed BB0056 data and gave an upper limit of the flux density $<0.8$\,mJy at 8.4\,GHz. 
In the S-band (2.3\,GHz), \obj\ is only marginally detected with a very low signal-to-noise ratio of 6.17 and a peak flux density of $0.82\pm0.24$\,mJy/beam. 

\subsection{Archival VLA Data}

To compare the radio morphology of \obj\ on different scales, 
we also retrieve two sets of archival data 
observed on 1992 December 31 using the VLA A-array configuration in the L-band and X-band (AM0384). 
The L-band data were taken at a central frequency of 1.4\,GHz and the X-band data were taken at a central frequency of 8.4\,GHz. 
Both observations have a total bandwidth of 100\,MHz. 
The data are manually calibrated using the Common Astronomy Software Application \citep[{\sc{casa}} v5.1.1;][]{2007ASPC..376..127M} following the standard procedures described in {\sc{casa}} Cookbook. 
The final images are made using the {\sc{difmap}} software package. 
The detailed description of the data reduction and analysis can be found in \citet{2020ApJ...904..200Y}.

\section{Results} \label{sec:results}

\begin{figure*}
    \centering
    \includegraphics[width=0.98\textwidth]{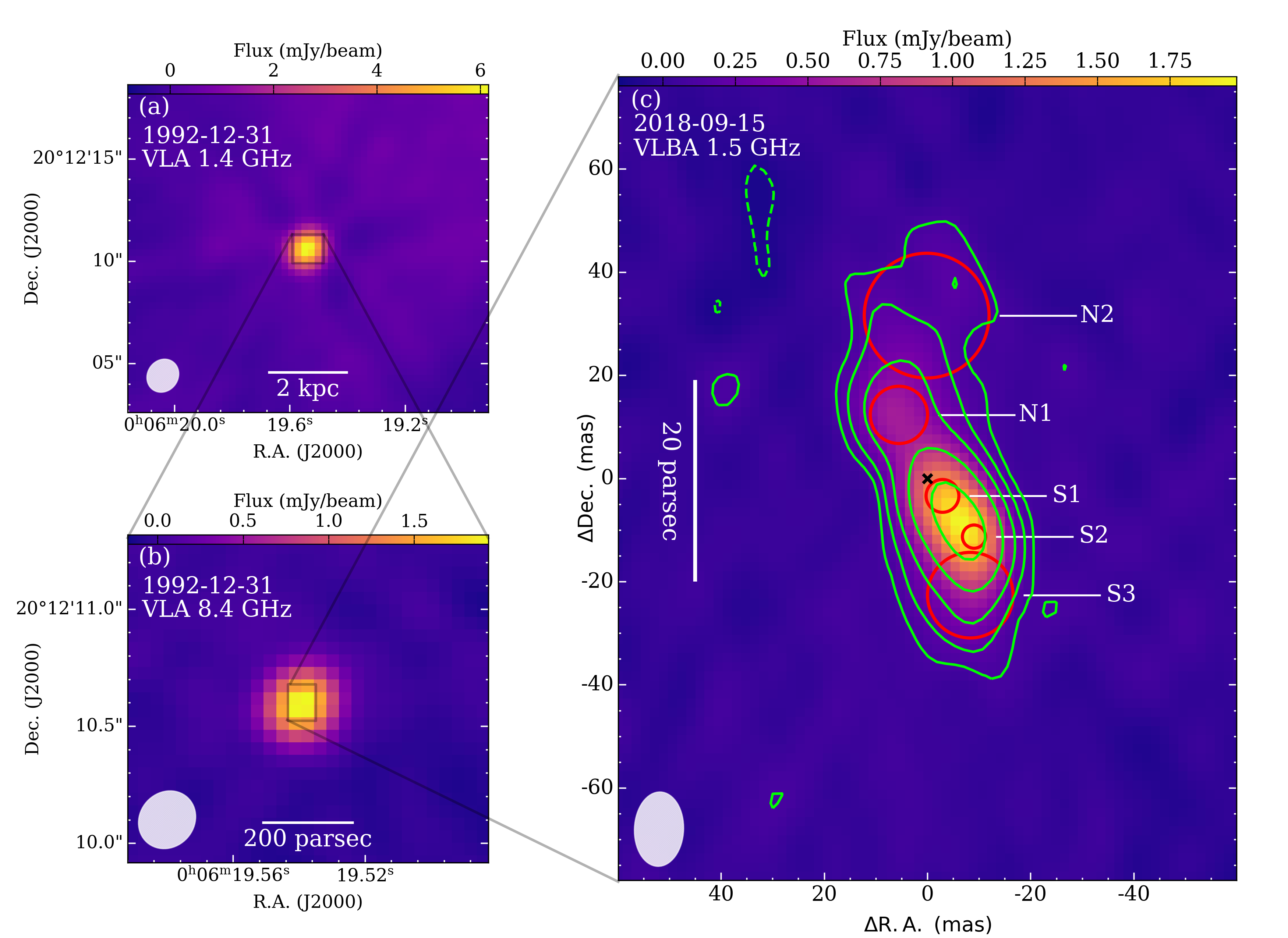}
    \caption{
    {\bf(a)} The VLA 1.4\,GHz image of \obj. 
    {\bf(b)} The VLA 8.4\,GHz image of \obj. 
    {\bf(c)} The VLBA 1.5\,GHz image of \obj. 
    The green contours in the VLBA image are plotted as 3$\sigma\times(-1, 1, 2, 4, 8, 16)$, 
    where $\sigma=0.03$\,mJy/beam is the rms noise. 
    The peak flux density is $1.98\pm0.05$\,mJy/beam. 
    The elongated jet structure is fitted using the two-dimensional Gaussian models. 
    The red circles
    mark the locations and sizes of the best-fit Gaussian components. 
    The optical position of \obj\ obtained by \gaia\ is represented by the black cross. 
    The synthesized beam size is illustrated as an ellipse at the lower left corner of each image. 
    \label{fig:radio_map}}
\end{figure*}


\subsection{Source Structures}

\begin{figure}
    \centering
    \includegraphics[width=0.9\columnwidth]{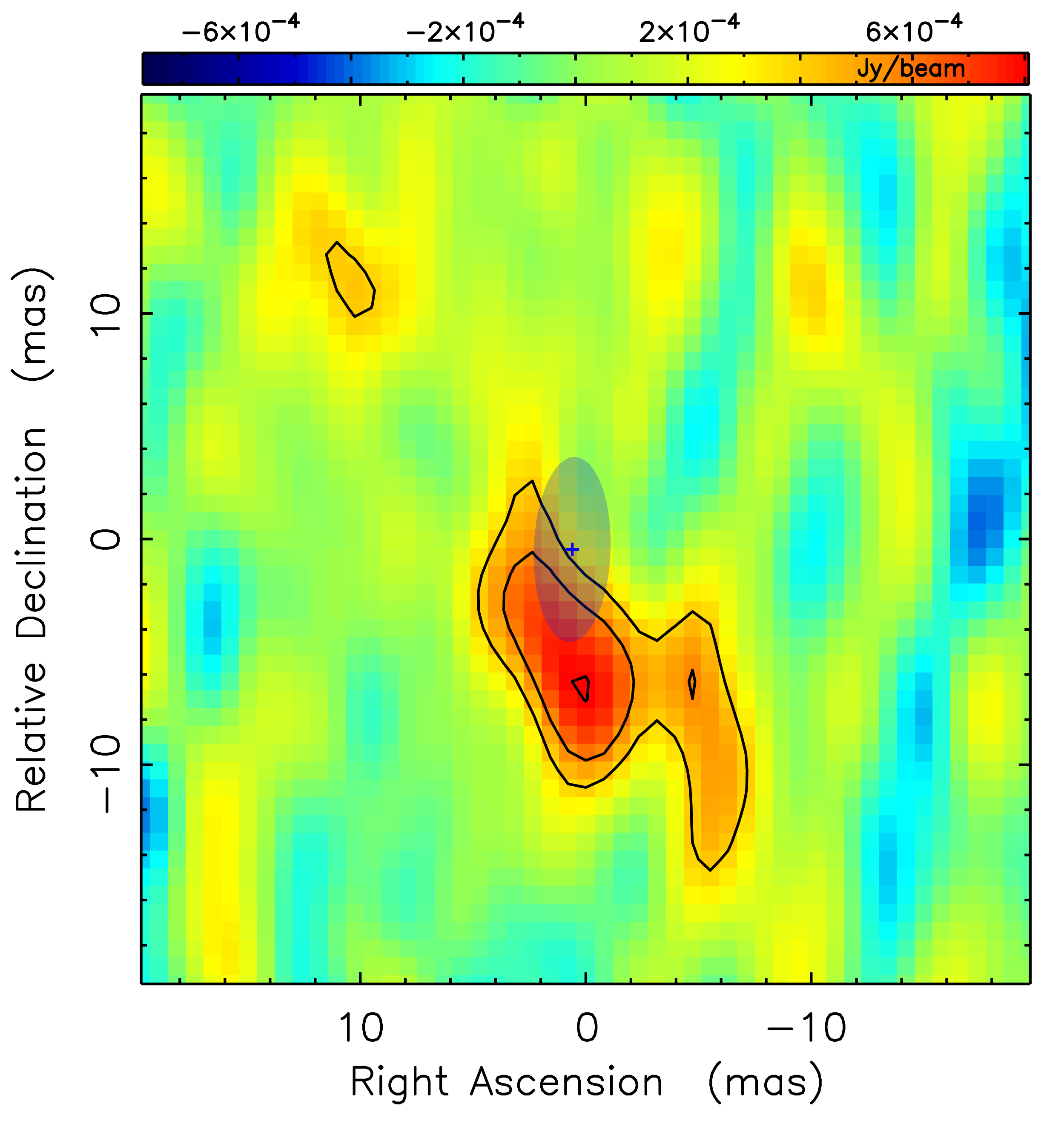}
    \caption{
    The archival VLBA S-band (2.3\,GHz) image of Mrk 335 (project ID: BG0093). 
    The image was produced by using a natural weight. 
    The contours are plotted as $3\sigma\times(-1,~1,~1.41,~2,~2.82)$, where $\sigma=0.13$\,mJy/beam is the rms noise. 
    The blue cross marks the location of the \gaia\ detection of \obj. 
    The beam is centred on \gaia\ position to show a VLBI positional uncertainty, 
    with its size of $8.31\times3.47$\,mas and a position angle of $-2.81^\circ$.
    \label{fig:archivevlbi}}
\end{figure}

The naturally weighted images of \obj\ observed by the VLA A-array configurature at 1.4\,GHz and 8.4\,GHz, and by the VLBA at 1.5\,GHz are shown in Figure~\ref{fig:radio_map}a, \ref{fig:radio_map}b and \ref{fig:radio_map}c, respectively. 
As can be seen, 
while \obj\ is compact, unresolved in the VLA images, 
it reveals 
an elongated structure extending 
along the south-north direction 
in the VLBA image. 
The structure is resolved at $\sim40$\,mas, corresponding to a projected physical scale of $\sim20$\,parsec at the redshift of \obj.
We also show the naturally weighted image of the archival S-band (2.3\,GHz) data in Figure~\ref{fig:archivevlbi}.
Due to the short exposure time and narrow bandwidth, 
the 2.3\,GHz image has a low signal-to-noise ratio 
and only recovers the brightest component (southern structure) as seen in the 1.5\,GHz image. 
Thus, we only consider the structure revealed by the new VLBA data at 1.5\,GHz in the following analysis. 

In order to further explore the structure seen in 1.5\,GHz VLBA image, 
we perform a two dimensional Gaussian model fitting to the visibility data 
using the {\tt{modelfit}} procedure in the {\sc{difmap}} package. 
The best-fit five components are listed in Table~\ref{tab:vlba_model}. 
The uncertainty of the flux density of each component is estimated by combining the fitting error with the initial calibrating error 
\citep[][]{2012AJ....144..105H}.
The entire structure of \obj\ has a total flux density of $6.46\pm0.06$\,mJy. 
The positions and sizes of the best-fit five components are displayed in Figure~\ref{fig:radio_map} as red circles, 
which likely reveals an S-shaped distribution.

\begin{table}
\setlength{\tabcolsep}{3pt}
\caption{Model-fitting results of 1.5\,GHz image obtained by our VLBA observation on \obj.}\label{tab:vlba_model}
\begin{center}             
\begin{tabular}{%
            c
            D{,}{\pm}{4.4}
            S
            S
            S
            S
            ccc}        
    \hline\hline                 
    \multicolumn{1}{c}{Comp.} & 
    \multicolumn{1}{c}{Flux Density} & 
    \multicolumn{1}{c}{Size}  &
    \multicolumn{1}{c}{$\Delta\alpha$} & 
    \multicolumn{1}{c}{$\Delta\delta$} & 
    \multicolumn{1}{c}{$\sigma_{\Delta}$} & 
    $T_\mathrm{B}$ 
    \\ &  
    \multicolumn{1}{c}{[mJy]} & 
    \multicolumn{1}{c}{[mas]} & 
    \multicolumn{1}{c}{[mas]} & 
    \multicolumn{1}{c}{[mas]} & 
    \multicolumn{1}{c}{[mas]} & 
    [K]  
    \\
    \multicolumn{1}{c}{(1)} &
    \multicolumn{1}{c}{(2)} &
    \multicolumn{1}{c}{(3)} &
    \multicolumn{1}{c}{(4)} &
    \multicolumn{1}{c}{(5)} &
    \multicolumn{1}{c}{(6)} &
    \multicolumn{1}{c}{(7)} &
    \\
    \hline
    N2    & 0.97, 0.01 & 24.18 & 0.14 & 31.55 & 0.38 & $1.3\times10^{6}$  \\
    N1    & 1.06, 0.01 & 11.10 & 5.51 & 12.31 & 0.16 & $6.6\times10^{6}$  \\
    S1    & 1.77, 0.02 &  6.35  & -2.93 & -3.40 & 0.05 & $3.3\times10^{7}$  \\
    S2    & 1.64, 0.02 &  4.50  & -9.04 & -11.29 & 0.04 & $6.2\times10^{7}$  \\
    S3    & 1.03, 0.01 & 16.59 & -8.32 & -22.64 & 0.25 & $2.8\times10^{6}$  \\
    \hline 
\end{tabular}
\parbox[]{\columnwidth}{%
    {\bf Note. }
    Column 1: labels of the components from north to south in Figure~\ref{fig:radio_map}c; 
    Column 2: 1.5\,GHz flux density and its uncertainty of the component; 
    Column 3: full-width-half-maximum of the 2 dimensional Gaussian model fitted to the component; 
    Column 4 and 5: the offset between the position of the radio component and the \gaia\ optical position in right ascension and declination, respectively; 
    Column 6: uncertainty of the offset; 
    Column 5: the radio brightness temperature calculated using Eq.~\ref{eq:bt}.
    }
\end{center}
\end{table}

The \gaia\ mission \citep[][]{2016A&A...595A...1G} has reported an optical position of \obj\ as 
RA=00$^\mathrm{h}$06$^\mathrm{m}$19$^\mathrm{s}$.537304$\pm$0$^\mathrm{s}$.00001,
Dec=$+$20$^\circ$12$^\prime$10$\farcs$61670$\pm$0$\farcs$0002 
(black cross in Figure~\ref{fig:radio_map}c). 
As can be seen in the figure, 
even considering the positional uncertainties of \gaia\ and the radio components,
there is still a significant offset between the optical position and the centre of nearest radio component. 
With \gaia\ position as reference,
the north and south radio components are respectively labelled 
as N2, N1 and S1, S2, S3. 
The symmetric distribution indicates a bipolar structure, 
with `S' (bright) components being the approaching jet boosting radio emission 
and `N' (faint) components being the receding jet fading away.

\subsection{Radio Brightness Temperature}

We calculate the radio brightness temperature for each model component from 
\begin{equation}\label{eq:bt}
T_{\rm B}=
1.8\times10^{9}(1+z)
\frac{S_{\nu}}{\nu^{2}\phi^{2}}
\end{equation}
in Kelvin \citep[K,][]{2005ApJ...621..123U}  in the rest frame of \obj, 
where $z$ is the redshift, 
$S_{\nu}$ is the integrated flux density in mJy at the observing frequency $\nu$ in GHz, 
and $\phi$ is the fitted FWHM of the Gaussian components in units of mas. 
All the Gaussian components have the radio brightness temperature 
$T_{\rm B}>10^{6}$\,K 
and the component ``S2'' has the highest brightness temperature of 
$T_{\rm B}\approx6.2\times10^{7}$\,K 
(Table~\ref{tab:vlba_model}).




\section{Discussion} \label{sec:discussion}

At parsec scales, 
we have resolved \obj\ at 1.5\,GHz into an elongated structure. 
The two dimensional Gaussian model fitting to the visibility data has revealed five components distributed along the elongated structure, 
among which ``S1'' and ``S2'' components have $T_{\rm B}>10^{7}$\,K, 
indicating the non-thermal emission from jet. 
This provides a strong evidence for a parsec-scale jet launched by the SMBH in \obj.
As mentioned in Section~\ref{sec:intro}, 
although the detailed physical mechanism for launching a jet is still elusive, 
it is generally suggested that the jet is coupled with the accretion process. 
Any detection of jet features in an object with accretion rate close to or above the Eddington limit would be of high value for the investigation of jet-accretion disk coupling. 
In the following we discuss the Eddington ratio of \obj, the origin of its radio emission, 
and the implication for its jet formation based on our detection.

\subsection{Eddington ratio of \obj}

%
Based on the reverberation mapping technique, 
the estimations of the black hole mass of \obj\ ranges from 
$8.3\times10^{6}\rm\,M_{\odot}$ to 
$2.6\times10^{7}\rm\,M_{\odot}$ 
\citep[][]{2004ApJ...613..682P, 2012ApJ...744L...4G, 2014ApJ...782...45D}, 
much lower than those in the classical blazars and radio galaxies \citep[e.g.][]{2002ApJ...569L..35F, 2003ApJ...583..134B}. 
Besides the black hole mass, 
the determination of Eddington ratio requires an estimate of the bolometric luminosity $L_{\rm bol}$ 
which is the integration of thermal emission released by the accretion process.
%
A direct 
way to estimate $L_{\rm bol}$ is to integrate the luminosity over a set of broad bands dominated by the emission from accretion disk. 
\citet{1994ApJS...95....1E} have constructed a spectral energy distribution (SED) from radio to X-rays of \obj\ 
and calculated the integral luminosity as $\sim3\times10^{45}$\,\lum\ 
by simply interpolating through the observed data, 
which would yield an Eddington ratio of $\lambda_{\rm Edd}\approx0.9$--$3$ considering the black hole mass range from the reverberation mapping. 
\citet{2009MNRAS.392.1124V} have built two sets of simultaneous SEDs from optical to X-rays of \obj, 
one of which was in its historical low X-ray flux state \citep[][]{2008ApJ...681..982G}. 
They calculated the integral bolometric luminosities based on the accretion disk (optical/UV)  and the power-law (X-ray) model fitted to the data. 
The bolometric luminosities are $2\times10^{45}$\,\lum\ and $1\times10^{45}$\,\lum, respectively, 
at those two epochs, which would yield an Eddington ratio of $\lambda_{\rm Edd}\approx0.3$--$1.9$.

There are several main factors contributing to the uncertainty of $L_{\rm bol}$. 
One is the lack of extreme ultraviolet (EUV) data, where the emission from \obj's accretion disk is supposed to peak. 
While the simple interpolation of the EUV gap between optical and X-rays may lead to an underestimation of $L_{\rm bol}$, 
the physically reasonable estimate of EUV peak is model-dependent. 
Another one which could lead to the underestimation of $L_{\rm bol}$ is the internal extinction of the optical/ultraviolet emission.
On the other hand, 
the contamination from the host galaxy can lead to an overestimation, mainly at longer wavelength, of $L_{\rm bol}$.
In optical, 
the host galaxy contribution to the emission at 5100\,\AA\ was estimated as $\lesssim20\%$ from a decomposition of \obj\ \hst\ image 
\citep[][]{2009ApJ...697..160B, 2014ApJ...782...45D}. 
Finally, the variability can also lead to the uncertainties in $L_{\rm bol}$. 
The flux of \obj\ varied by only a few per cent in the optical and by $\sim20\%$ in the UV
\citep[][]{2014ApJ...782...45D, 2018MNRAS.478.2557G}. 
Meanwhile it has exhibited much higher variability in X-rays by a factor of $\sim50$ 
\citep[e.g.][]{2018MNRAS.478.2557G, 2020MNRAS.499.1266T}. 
However, 
since the coronal optical emission lines, 
which are supposed to be driven by EUV and soft X-rays, 
do not vary much in \obj\ \citep[][]{2008ApJ...681..982G}, 
it was suggested that such huge X-ray variability is due to the variable column density and covering factor of the X-ray absorbers near the SMBH \citep[][]{2020A&A...643L...7K}. 
Therefore, even considering the above uncertainties of $L_{\rm bol}$, 
the Eddington ratio of \obj\ is still high.

At last, we should note that, 
according to the current understanding of the accretion physics, 
the `slim disk' is used to describe the accretion process around SMBH when Eddington ratio approaches unity \citep[][]{1988ApJ...332..646A, 2000PASJ...52..499M}. 
The optical depth of such slim disk is so large that the majority of photons will be trapped in the disk until advected
into the black hole rather than being radiated out, 
significantly reducing the radiation efficiency. 
It means that the actual mass accretion rate will be even higher than predicted from the luminosity assuming typical radiation efficiencies in a standard optically thick, geometrically thin disk \citep[][]{2014ApJ...793..108W}. 
Thus, the SMBH in \obj\ should reliably be a highly accreting black hole.

\subsection{Origin of Radio Emission}

\obj's radio emission is relatively weak. 
At 1.4\,GHz, \citet{1976A&A....53...93D} have observed \obj\ 
with the Westerbork telescope but were only able to set a 3$\sigma$ upper limit of $f_{1.4}<8$\,mJy. 
\citet{1995ApJS...98..369B} have observed \obj\ with the {\it Arecibo} telescope and reported an upper limit of $f_{1.4}<5$\,mJy. 
The observation at 2.3\,GHz performed by the Parkes-Tidbinbilla Interferometer with a sub-arcsec resolution gives a flux density of $f_{2.3}=5$\,mJy \citep{1994ApJ...432..496R}. 
The observation at 20\,GHz with the Owens Valley Radio Observatory reveals an upper limit of $f_{20}<1.7$\,mJy \citep{1987ApJ...313..651E}. 

During the past decades, \obj\ has also been observed by the VLA with various angular resolutions (Table~\ref{tab:vla_flux}). 
The angular resolution of these observations vary from 0$\farcs$3 to 1$\farcm$5, 
corresponding to a linear size from 150\,parsec to 46\,kpc. 
Only a compact unresolved source was found in \obj\ in all of these observations, 
and no extended emission has been detected. 
At 1.4\,GHz and 5\,GHz, \obj\ shows slight variability. 
At 8.4\,GHz, \citet{2005ApJ...618..108B} have carried out several epochs of the VLA observations between 1997 January and 1999 April, 
but didn't find significant variability. 
Thus, the radio emission of \obj\ is compact and rather stable, 
with slight variability at lower frequencies. 
We estimate the radio spectral index between 1.4\,GHz and 8.4\,GHz using the simultaneous flux measurements on February 1992 (Table~\ref{tab:vla_flux}) and obtain $\alpha_{\rm r}\approx-0.6$. 
The spectral index between 5\,GHz and 8.4\,GHz is estimate as $\alpha_{\rm r}\approx-0.8$ to $-1.0$ using the flux measured at the angular size of $0\farcs3$ (Table~\ref{tab:vla_flux}). 
If we consider the peak flux densities measured by the VLBA observations in the L-band (1.5\,GHz) and S-band (2.3\,GHz) (Section~\ref{sec:vlba_observation}), 
an even steeper spectral index of $\alpha_{\rm r}\sim-2$ is obtained, 
although the uncertainty of this index would be large due to the possible flux variability at different epochs and the low signal-to-noise ratio of the S-band image. 
While the GHz emission from the star-forming regions is usually characterized by steep spectra \cite[][]{1992ARA&A..30..575C}, 
the steep spectral radio emission was also observed from nuclear outflows and jets 
\citep[e.g.][]{2004ApJ...613..794G, 2012MNRAS.426..588B}. 
We address the contributions to the observed radio emission in \obj\ from star-forming activity, outflows and jets as follows.

\subsubsection{Star-forming activity}

It was suggested that the NLS1 galaxies typically have stronger star-forming activities compared to the normal Seyfert galaxies 
\citep[e.g.][]{2006AJ....132..321D, 2010MNRAS.403.1246S}. 
In this case, the star formation process in their dusty spirals or circumnuclear regions can have significant contributions to the radio emission. 
By analysing the mid-infrared (mid-IR) colors and the ratio of the radio to mid-IR flux for a sample of radio-loud NLS1 galaxies, 
\citet{2015MNRAS.451.1795C} found that the star-forming activities can contribute a significant fraction of radio emission even in some radio-loud NLS1 galaxies. 
But this is not likely the case in \obj. 
The overall radio morphology of the star-forming regions are typically extended, diffuse and clumpy \citep[e.g.][]{1987A&A...172...32H, 2010A&A...513A..11O}, 
while the radio images of \obj\ only reveal a compact source at various resolutions (e.g. Figure~\ref{fig:radio_map}a and \ref{fig:radio_map}b). 
The flux densities measured at scales from tens of kpc to a few tens of parsec are quite consistent, 
indicating that the star-forming activity makes very little contribution to the total radio emission. 
This is supported by the fact that \obj\ has a very low star formation rate (SFR) among NLS1 galaxies. 
\citet{2021ApJ...910..124X} have estimated a SFR of $\sim1\,M_{\odot}\rm\,yr^{-1}$ 
based on the infrared SED and spectroscopy, 
which would yield a radio flux density $<1$\,mJy at 1.5\,GHz following the empirical relation derived by \citet{1992ARA&A..30..575C}.


\subsubsection{Wind-driven outflows}

The accretion disk winds can drive wide-angle outflows, 
which are suggested to be prevalent in luminous AGN and are likely stronger in objects with higher Eddington ratios 
\citep[e.g.][]{2002ApJ...569..641L, 2007ApJ...665..990G, 2014ApJ...786...42Z}. 
In the nuclear region, 
these outflows have been typically observed as the blueshifted absorption lines in the UV/X-ray spectroscopy \citep[e.g.][]{2019ApJ...875..150L}. 
On the galactic scales, 
they have been detected as the blue-wing components in the narrow emission line profiles \citep[e.g.][]{2008ApJ...680..926K} or the molecular line profiles \citep[e.g.][]{2015A&A...583A..99F}. 
The wind-driven outflows may also produce radio emission 
via bremsstrahlung free-free process from ionized plasma 
\citep[][]{2007ApJ...668L.103B} 
or synchrotron process of the relativistic electrons accelerated by the wind shocks 
\citep[][]{2014MNRAS.442..784Z}, 
with the later being used to explain the correlation between radio emission and the [O{\sc\,iii}] line velocity width in the radio quiet AGN 
\citep[e.g.][]{2014MNRAS.442..784Z, 2018MNRAS.477..830H}. 
In the case of \obj, 
evidence for outflows with velocity of a few thousand \kmps\ is observed as blueshifted absorption lines in its X-ray and UV spectra 
\citep[e.g.][]{2013ApJ...766..104L, 2019ApJ...875..150L}. 
But, again, we note that the radio images of \obj\ do not reveal any extended diffused radio emission on scales from kpc all the way down to a few tens of parsec. 
Thus, if the wind-driven uncollimated outflows have made any detectable contribution to the radio emission, 
it would likely originate from the central parsec-scale region. 

\subsubsection{Parsec-scale bipolar jet structure}

On parsec scales, the  VLBA image at 1.5\,GHz shows a linear elongation in the north-south direction. 
The two dimensional model fitting to the visibility data has revealed components in the elongated structure with the brightness temperature of 
$T_{\rm B}>10^{7}$\,K 
(Table~\ref{tab:vlba_model}),
indicating a non-thermal, synchrotron origin. 
It strongly suggests that the radio emission detected on parsec scale is from the jet \citep{2019NatAs...3..387P}.
The offset between \gaia\ optical position and the nearest radio peak position is $\Delta\approx5$\,mas, 
corresponding to a projected linear offset of 2.5\,parsec.
Taking the uncertainties in both R.A. and Dec. from the \gaia\ measurement ($<0.1$\,mas) and the VLBA measurement (Table~\ref{tab:vlba_model}) into account, 
this offset is still significant.
The nuclear optical emission is usually from the accretion disk of the central engine.
If the position of the central engine is really offset from either of the radio peaks, and is located between `N' and `S' components,
considering the symmetric jet distribution with regards to the central engine represented by \gaia\ 
(Figure~\ref{fig:radio_map}), 
the radio structure is likely to indicate the bipolar ejection with a moderate angle between its axis and the line of sight.
The larger flux density of the south branch than the north one could be due to the beaming effect as the south jet is approaching the observers. 
And if the `S' components and `N' components represent where the jet and counterjet is located, respectively, 
we would derive a low jet-to-counterjet ratio of $\sim2.2$. 
A low jet-to-counterjet ratio usually results from either that the jet has a large viewing angle between its axis and the line of sight, e.g. close to the sky plane, 
or that the jet has a low velocity. 
Given that \obj\ is a type 1 AGN, the viewing angle is unlikely to be very large. 
So the low jet-to-counterjet ratio may imply a low jet speed in \obj. 
For instance, 
assuming a viewing angle of $45^{\circ}$ and a spectral index of $-0.5$, 
we obtain a jet speed of $\beta_{\rm j}=(v_{\rm j}/c)\approx0.2$ by using the equation in \citet[][]{2020MNRAS.496..676B}. 



The total flux density of the detected components in the VLBA image is $6.46\pm0.06$\,mJy, 
which is similar to 
the measurements on larger scales taken 
by the VLA with poorer resolution (Table~\ref{tab:vla_flux}). 
Especially, the total radio flux density detected in the VLBA L-band is consistent with the $1.4$\,GHz VLA flux density of $6.19\pm0.29$\,mJy measured with a resolution of $\sim$2\,arcsec \citep{2020ApJ...904..200Y} within uncertainty, 
indicating that the radio emission within 2\,arcsec (corresponding to $\sim1$\,kpc) is dominated 
by the central parsec-scale bipolar jet structure and that the contribution from the star-forming activity and wind-driven outflows is very small 
even if it can not be totally ruled out. 

We notice that the two dimensional Gaussian models fitted to the VLBA visibility data at $1.5$\,GHz reveal an S-shaped distribution, likely implying a helical jet or slowly drifting of the position angle of the jet axis.
One of the scenarios adopted to explain the change of the jet axis is 
the binary black hole system, 
in which the secondary black hole causes the precession of the black hole orbit or the primary accretion disc, 
resulting in an S-shaped jet 
\citep[e.g.][]{1993ApJ...409..130R, 2017MNRAS.465.4772R}.
But there is not yet evidence, e.g. double-peaked emission lines, systematic line shifts, periodic behaviour or dual bright compact radio cores, for the existence of binary black hole system in \obj. 
Another possible mechanism causing the S-shaped jet would be the warped disk near the jet launching site, 
arising from non-axisymmetric radiation \citep[][]{1996MNRAS.281..357P} 
or accretion events, 
the latter of which would be interesting given that \obj\ is accreting at such high Eddington ratio and has been detected for several flares in optical and X-rays during the past decades
\citep[][]{2020A&A...643L...7K}. 
However, the current data set is not adequate to confirm the validity of they S-shaped jet structure in \obj.
Further observations are required to investigate the outer structures at both jet sides with medium-resolution and the inner jet structures with high-resolution, 
which are crucial to confirm the S-shaped jet structure and its origin.


\subsection{Implications on Jet Formation}\label{sec:dis_imp}

The jets have been widely observed in different types of accreting sources. 
During the past decades, great efforts have been made on theoretical work to explain the jet formation. 
So far, several popular models have been proposed such as the Blandford--Znajek (BZ) process, 
in which the jet is driven by extracting the rotational energy of the black hole via a large-scale magnetic
field \citep[][]{1977MNRAS.179..433B}, 
and Blandford--Payne (BP) process, 
in which the jet is driven by extracting the rotational energy of the accretion disk via the magnetic fields threading the disk \citep[][]{1982MNRAS.199..883B}. 
Recently, some numerical simulations have been done for the jet formation in super-Eddington accreting supermassive black holes, 
in which the jet is driven by radiation pressure 
\citep[e.g.][]{2011ApJ...736....2O, 2014MNRAS.437.2744T, 2015MNRAS.453.3213S}. 
However, although a lot of important progress has been made, 
we still do not fully understand the physical mechanism for launching the jet.

\subsubsection{Analog to Galactic black holes?}

The radio structure observed in \obj\ by VLBA provides a good evidence of jet launched by a highly accreting SMBH in the radio-quiet Seyfert galaxy. 
Although the detailed mechanism of jet formation is still unknown, 
a possible way to explore the jet formation in 
\obj, 
in the framework of current knowledge, 
is through the comparison with other highly accreting sources. 
As the down-scaled analogues to the accreting SMBHs in AGN, 
the Galactic black holes (GBH) of stellar mass in the X-ray binaries also display the phenomenon of accretion (in X-rays) and jet (in radio). 
In the GBH, 
the steady jet is always associated with low-luminosity hard X-ray spectral state (low/hard state), 
while the radio emission is suppressed in the high-luminosity soft X-ray spectral state (high/soft state). 
In addition, 
an intermediate state (i.e. very high state) is also observed when GBH's soft X-ray luminosity rises to a peak and 
the jet becomes choked and episodic, 
associated with the radio outburst 
\citep{2004MNRAS.355.1105F}. 

Considering the high Eddington ratio, 
\obj\ may be the scaled-up version of GBHs at the end of their very high state. 
If this is the case, the radio observations may capture the evidence of its past ejecting activities during the state transition. 
Previous observations have found several NLS1 galaxies showing kiloparsec-scale radio structures, 
implying the past ejecting activities \citep[e.g.][]{2010ApJ...717.1243G, 2012ApJ...760...41D, 2015ApJ...800L...8R, 2020ApJ...904..200Y}. 
However, such evidence has not yet been detected in \obj\ by observations spanning the past decades. 
Alternatively, \obj\ may have already transitioned to the high/soft state so long ago 
that the energy of the ejecta, if it had existed, has already dissipated, 
therefore
kiloparsec-scale radio emission tracing the past ejecting activities is not detected.
If this is the case, 
the elongated structure detected by our VLBA image should be newly ejected after the steady/episodic jet being switched off, 
which implies that the collimation of the outflows in \obj\ could be maintained/established on small scales 
even after the accreting source has transitioned to the high/soft state.

\subsubsection{Comparison to tidal disruption event}

Another kind of highly accreting source is the tidal disruption event 
\citep[TDE;][]{2015JHEAp...7..148K}, 
in which a star passing by a SMBH is tidally disrupted and accreted by the SMBH, 
leading to a bright flare with an accretion rate near or exceeding the Eddington limit. 
Several dozens of TDEs have been discovered based on their large-amplitude brightening in X-rays, UV and/or optical. 
But so far only three TDEs have been reported to launch jets 
\citep[][]{2011Sci...333..203B, 2011Natur.476..421B, 2011Sci...333..199L, 2011Natur.476..425Z, 2012ApJ...753...77C, 2016ApJ...819L..25A, 2016Sci...351...62V}. 
For instance, as the most representative jetted--TDE, 
Swift~J1644+57, 
which hosts an inactive galactic nuclei, 
has revealed an apparent luminosity $\sim10^{3}$ times of the Eddington limit for its black hole mass, 
and a spectral energy distribution characterized by synchrotron and inverse Compton processes 
\citep[][]{2011Sci...333..203B, 2011Natur.476..421B, 2011Sci...333..199L, 2011Natur.476..425Z}, 
leading inevitably to a scenario of a jet pointing to the observer. 
A variable radio synchrotron emission was detected 
a few days after the rapid rise in its X-rays \citep[][]{2011Natur.476..425Z}, 
providing evidence for jet launched by the highly accreting SMBH. 
Recent numerical simulations of the super-Eddington compact disk used to explain the TDEs 
have shown that, together with a geometrically thick disk and a fast outflow, 
a relativistic jet can also be produced 
under optimal conditions 
given a spinning black hole and a large-scale ordered magnetic field 
\citep[][]{2018ApJ...859L..20D}. 
If the accretion process in the highly accreting AGN 
is the same as that in the TDE, 
this gives hints to the jet feature in \obj: 
the jet may be driven by the magnetic field threading the spinning SMBH and the disk, 
and is powered by the black hole spin energy. 
This is consistent with a high spin parameter measured for \obj\ based on the blurred reflection model fitted to the X-ray spectrum 
\citep[][]{2015MNRAS.446..633G, 2019MNRAS.484.4287G}. 
The size of the jet in \obj\ is very small and only resolved at parsec scales (Figure~\ref{fig:radio_map}). 
Coincidently, the radio structure in jetted-TDE was also found to be very compact 
\citep[][]{2016MNRAS.462L..66Y, 2016ApJ...832L..10R}. 
Considering the low jet speed and the high Eddington ratio in \obj, 
the small size of the jet may be due to the jet deceleration by interaction with surrounding dense medium in the nuclear region. 



\subsubsection{Missing radio core?}

In a jet structure, a parsec-scale compact radio core with flat or inverted spectrum is usually present at the position of jet base in the vicinity of accreting SMBH. 
As shown in Figure~\ref{fig:radio_map}, 
even taking into account the uncertainties, 
the optical position of \obj\ measured by \gaia\ does not coincide with the position of any radio components. 
If the optical position represents the position of the central SMBH and its accretion disk, 
our results imply an offset between the radio components and the central engine in \obj, 
and we do not explicitly find the radio core near the position of the central engine. 

The symmetric jet structure lacking a detectable core was also reported in other kind of radio sources such as some compact symmetric objects, 
in which the core is so weak due to the jet axis very close to the sky plane \citep[e.g.][]{2000ApJ...541..112T}. 
But this is unlikely the case in \obj\ as it is a type 1 AGN. 
We note that the emission centre in optical and radio may not always coincide due to that the centroid of the radio component may be shifted with observing frequency 
\cite[][]{2008A&A...483..759K, 2017MNRAS.467L..71P}. 
So a possible reason for the `missing core' in our observation may be that the core is too weak to be detected at 1.5\,GHz due to the synchrotron-self-absorption at lower frequencies 
\citep[][]{2017A&A...598L...1K}, 
and it is buried in other radio components and not resolved by our observation. 
If this is the case, 
the core may 
reveal itself  at higher frequencies. 
Unfortunately, the archival VLBA observation of \obj\ at 8.4\,GHz (project ID: BB0056 and BG0093) does not reveal any detection, 
and only gives an upper limit of $<0.8\rm\,mJy$ 
\citep[][]{1998MNRAS.299..165B}. 
Deeper VLBI observations at higher frequencies, e.g. 8.4\,GHz and 15\,GHz, 
are needed to test if there exists a compact core. 

\subsubsection{Jet-corona connection}

A collimated corona has been suggested to launch from the accretion disk based on the past X-ray spectroscopic observations on \obj\ during its X-ray flares 
\citep[e.g.][]{2015MNRAS.449..129W, 2019MNRAS.484.4287G}. 
Generally, the X-ray emission is produced by the hot electrons in corona above the accretion disk via inverse Compton scattering of the soft photons from disk. 
Then a fraction of X-ray emission illuminates the disk and is reflected, 
contributing to the reflection component in the observed X-ray spectrum. 
During the X-ray flares of \obj, 
it was found that the reflection fraction dropped significantly and the emissivity profile of the outer disk steepened, 
indicating that the corona is illuminating less to the disk, 
which is expected if the corona is moving away from the disk at high speed 
and beaming its radiation 
\citep[][]{2019MNRAS.484.4287G}. 
In addition, 
it was suggested that the AGN corona might be magnetically heated, 
and the radio and X-ray emission originating from the coronal activities would reveal a ratio of 
$\sim10^{-5}$, 
similar to those in the coronally active stars 
\citep[][]{2008MNRAS.390..847L}.
Using the X-ray fluxes provided in \citet{2008ApJ...681..982G} and the radio measurements in Table~\ref{tab:vla_flux} as well as by our VLBA measurement, 
the flux ratio of radio-to-X-ray is in the range $\sim10^{-6}$ to $\sim10^{-5}$ 
which is similar to the suggested relation in \citet{2008MNRAS.390..847L}. 
It implies the possibility that the corona is the jet base, 
which was also suggested by previous works, e.g., in \citet{2015MNRAS.449..129W} and \citet{2019MNRAS.484.4287G}.

However, 
the ejecta in the collimated corona detected by the X-ray can only reach a height of no more than 
ten gravitational radius $R_{\rm g}$ as estimated in \citet{2019MNRAS.484.4287G} (see their Figure 10),
while the projected scale of the jet structure detected in our VLBA image is $\sim20$\,parsec, 
corresponding to $\sim10^{7}\,R_{\rm g}$. 
It is not clear whether the parsec-scale radio jet in \obj\ has any relation to the collimated X-ray corona, 
and, if they are related, how the particles can be accelerated to such large distance from within a few $R_{\rm g}$ above the accretion disk. 
This may be tested by the future higher-resolution radio observations and by the correlation of simultaneous X-ray and radio monitoring observations. 

\section{Summary}

We report the VLBA observation at 1.5\,GHz on a nearby radio-quiet NLS1 galaxy \obj\ in which its central SMBH accretes near/above Eddington limit. 
The image reveals an elongated jet structure extending $\sim20\rm\,parsec$, 
with the brightness temperature as high as $\sim6\times10^{7}\rm\,K$. 
This work increases the number of known parsec-scale jet in the highly accreting radio-quiet NLS1 galaxies, which is rare so far. 
The consistence of the flux measured by VLBA with the historical measurements at larger scales by VLA shows that the radio emission of \obj\ is dominated by the central parsec-scale jet. 
We discuss the comparison with other highly accreting jetted black holes, and the possible connection between the X-ray corona and the jet. 
The discovery of jet features in the highly accreting AGN will provide us with further constraints on the physics of the accretion-jet coupling in high Eddington ratio regime. 
In our future work, we will report results of the observations on the parse-scale nuclear region in other highly accreting radio-quiet NLS1 galaxies (Yang et al. in preparation). 

\section*{Acknowledgements}

SY acknowledges the support by an Alexander von Humboldt Foundation Fellowship. 
XLY thanks the support by Shanghai Sailing Program (21YF1455300) and China Postdoctoral Science Foundation (2021M693267). 
XLY and TA thank the financial support by 
the National Key R\&D Programme of China (2018YFA0404603) and 
the Chinese Academy of Sciences (114231KYSB20170003). 
MFG is supported by the National Science Foundation of China (11873073). 
LCH, RW and XW was supported by the National Science Foundation of China (11721303, 11991052) and the National Key R\&D Program of China (2016YFA0400702 and 2016YFA0400703). 
The National Radio Astronomy Observatory is a facility of the National Science Foundation operated under cooperative agreement by Associated Universities, Inc. 
This work made use of the Swinburne University of Technology software correlator, developed as part of the Australian Major National Research Facilities Programme and operated under licence. 
This work has also made use of the NASA Astrophysics Data System Abstract Service (ADS), and the NASA/IPAC Extragalactic Database (NED) which is operated by the Jet Propulsion Laboratory, California Institute of Technology, under contract with the National Aeronautics and Space Administration.

\section*{Data Availability}

The VLBA data of BY0145 will be shared on reasonable request to the corresponding author. 
The archival VLBA data (BB0056, BG0093) and the VLA data (AM0384) underlying this article are available in the NRAO data archive (\url{https://science.nrao.edu/facilities/vla/archive/index}).




\bibliographystyle{mnras}
\bibliography{references} 








\bsp	
\label{lastpage}
\end{document}